\documentclass[aps,prb,twocolumn]{revtex4-1}
\usepackage{graphicx}
\usepackage{amssymb}
\usepackage{amsmath}
\usepackage{enumitem}

\usepackage{color}
\usepackage[normalem]{ulem}

\begin{document}

\title{Interplay between nanometer-scale strain variations and externally \\ applied strain in graphene}

\author{G. J. Verbiest}
\email{verbiest@physik.rwth-aachen.de}
\affiliation{JARA-FIT and 2nd Institute of Physics, RWTH Aachen University, 52074 Aachen, Germany}
\author{S. E. Huber}
\affiliation{Chair for Theoretical Chemistry and Catalysis Research Center, Technische Universit{\"a}t M{\"u}nchen,\\ Lichtenbergstr. 4, 85747 Garching, Germany}
\author{M. Andersen}
\affiliation{Chair for Theoretical Chemistry and Catalysis Research Center, Technische Universit{\"a}t M{\"u}nchen,\\ Lichtenbergstr. 4, 85747 Garching, Germany}
\author{C. Stampfer}
\affiliation{JARA-FIT and 2nd Institute of Physics, RWTH Aachen University, 52074 Aachen, Germany}
\author{K. Reuter}
\affiliation{Chair for Theoretical Chemistry and Catalysis Research Center, Technische Universit{\"a}t M{\"u}nchen,\\ Lichtenbergstr. 4, 85747 Garching, Germany}

\begin{abstract}
We present a molecular modeling study analyzing nanometer-scale strain variations in graphene as a function of externally applied tensile strain. We consider two different mechanisms that could underlie nanometer-scale strain variations: static perturbations from lattice imperfections of an underlying substrate and thermal fluctuations. For both cases we observe a decrease in the out-of-plane atomic displacements with increasing strain, which is accompanied by an increase in the in-plane displacements. Reflecting the non-linear elastic properties of graphene, both trends together yield a non-monotonic variation of the total displacements with increasing tensile strain. This variation allows to test the role of nanometer-scale strain variations in limiting the carrier mobility of high-quality graphene samples.
\end{abstract}

\maketitle

\section{Introduction}

Graphene is a promising material with many remarkable properties.\cite{Lee,Nair,Castro,Trauzettel,Fratini,Katsnelson} However, its experimentally measured characteristics often do not match those theoretically predicted.\cite{Drogeler,Zhang} Prominently, for example, the carrier mobility of graphene samples was found to strongly depend on the fabrication process and type of substrate. On rough SiO$_2$ the mobility of graphene samples is e.g. limited to a few ten thousand cm$^2$/Vs,\cite{Zhang2,Zhang3} while using atomically flat hexagonal boron nitride (hBN) as a substrate
%for graphene devices was instead found to increase the
allows mobilities up to several hundred thousands cm$^2$/Vs or more at low temperatures.\cite{Dean,Abanin,Wang,Banszerus,Banszerus2} Notwithstanding, world-record mobility graphene devices consist of a suspended graphene sheet and can even reach one million cm$^2$/Vs.\cite{Bolotin,Du,Castro,Mayorov} As suspended graphene devices are not very practical, it is thus imperative to understand and tailor the carrier mobility limitations in supported graphene.
A key element for this will be %a detailed understanding of
the control over
the substrate induced
interplay between structural and electronic properties.
The truly two-dimensional nature of graphene makes this interplay, or generally the electromechanical coupling, very unique
and promises tailored electronic properties by so-called strain engineering.\cite{Hasegawa,Pereira3,Pellegrino,Cocco,Choi,Choi2,Leite,Schneider,Pereira2,Low,Amorim,Verbiest}

%as well as intrinsic electromechanical coupling in graphene important.
%One particularly interesting method is to utilize the electromechanical coupling in graphene to engineer the required electronic properties.\cite{Hasegawa,Pereira3,Pellegrino,Cocco,Choi,Choi2,Leite,Schneider} Within the frame of this so-called strain engineering, graphene is strained on a 'global' scale.

%DO NOT FORGET:
%This means that the entire flake, or a significant fraction of it, is under constant tension.

% GÓES TO SUMMARY:
%Recently, micro-electromechanical systems (MEMS) were introduced in the research field of graphene to do just that.\cite{Garza,Garza2,Zhang4}
%This allows for the experimental study of the interplay between nanometer-scale strain variations and externally applied strain, which is currently rather unknown territory.

%The effect of strain on the carriers in graphene is generally twofold. First, strain generates a potential that is similar to the vector potential of a real magnetic field and is therefore called the pseudovector potential.\cite{Suzuura,Manes,Sasaki,Katsnelson2,Juan,Masir,Adebpour,Kim} Second, the area of the unit cell is altered which results in a redistribution of the charge carrier density in a way that strain variations give rise to effective electron-hole puddles at low carrier densities.\cite{Suzuura,Manes}

The effect of strain on the carriers in graphene is generally twofold. First, the area of the unit cell is altered resulting in a redistribution of the charge carrier density in a way that strain variations give rise to effective electron-hole puddles at low carrier densities.\cite{Suzuura,Manes} Second, strain generates a so-called pseudovector potential that is similar to the vector potential of a real magnetic field.\cite{Suzuura,Manes,Sasaki,Katsnelson2,Juan,Masir,Adebpour,Kim,Burgos}
%An externally applied gate potential can thus never be used to reach exactly zero charge carrier density in graphene with
 %
It is exactly such a strain-induced randomly varying pseudovector potential, which
recently has been identified as the limiting mechanism for the
carrier mobility in high-quality graphene devices on a substrate.\cite{Couto,Engels,Neumann}
As Couto and co-workers\cite{Couto} showed, these strain variations are on a length scale of a few nanometers such that they act as long-range scattering centers allowing for pseudospin flips. Thus nanometer-scale
strain variations enable direct backscattering that limits the carrier mobility in high-qualtiy graphene samples.
In fact, it has been shown that the mobility is inversely proportional to the strength of nanometer-scale strain variations.\cite{Couto}
As these strain variations can be non-invasively monitored by the line-width of the graphene specific Raman 2D-line\cite{Neumann}, fabrication processes have been recently optimized to suppress strain variations leading  to significant improvements in carrier mobility.\cite{Banszerus,Banszerus2}
For future optimization a thoroughly understanding of the interplay between strain variations and external mechanical influences,
such as `global' strain are of great importance.
Also in view of the promise to tailor electronic properties by strain engineering a detailed knowledge of this particular interplay is crucial.

\begin{figure}[!t]
 \begin{center}
	\includegraphics[width=86mm]{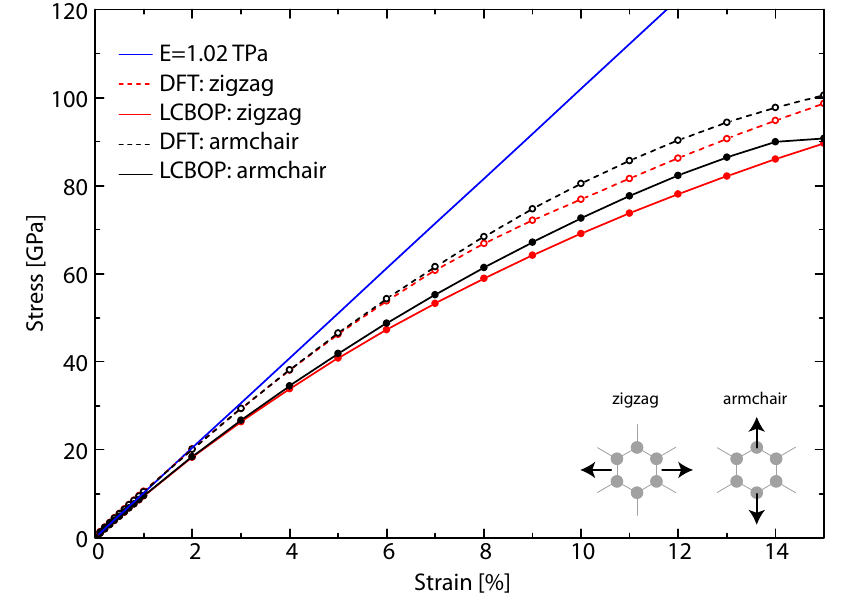}
 \end{center}
 \caption{(Color online) Stress-strain curves for a graphene sheet under tensile external strain up to 15\%. The linear behavior expected for a material with constant Young's modulus of 1.02 TPa \cite{Lee} (blue curve) is compared to calculated data for strain applied along the zigzag (red curves) and along the armchair (black curves) directions. The data obtained with the empirical LCBOP potential (solid lines) matches the one obtained by first-principles DFT calculations (dotted lines) within 10\%.}
 %%KR Please create this new figure from the lower panel of present Fig. 2. - GV: Done.
 \label{fig1new}
\end{figure}

%At present, not much is known about the properties and origin of these nanometer-scale strain variations.

%Despite that graphene is the material with the highest reported room temperature mobility, graphene has a problem for many practical applications: it is a semimetal\cite{Katsnelson}. Much research is focused on overcoming or finding a way to circumvent this problem.\cite{Pereira2,Low,Amorim} One particularly interesting method is to utilize the electromechanical coupling in graphene to engineer the required electronic properties.\cite{Hasegawa,Pereira3,Pellegrino,Cocco,Choi,Choi2,Leite,Schneider} Within the frame of this so-called strain engineering, graphene is strained on a 'global' scale. This means that the entire flake, or a significant fraction of it, is under constant tension. Recently, micro-electromechanical systems (MEMS) were introduced in the research field of graphene to do just that.\cite{Garza,Garza2,Zhang4} This allows for the experimental study of the interplay between nanometer-scale strain variations and externally applied strain, which is currently rather unknown territory.

In this work we discuss a molecular modeling study analyzing this interplay.
Under compressive strain nanometer-scale strain variations may be intricately mingled with the formation of static ripples.\cite{Castro,Meyer,Fasolino,Braghin,Monteverde} We therefore focus our analysis on applied tensile strain.
As for the physical origin of the nanometer-scale strain variations in graphene, thermal fluctuations as well as frozen ripples from the fabrication process and atomic defects in substrates like hBN are conceivable\cite{Castro,Mariani,Morozov,Stauber,Anota,Park,Satos}, we consider both the effect of thermal fluctuations and of a static Gaussian potential to model defects in a hBN substrate.
%
%Both thermal fluctuations and frozen ripples from the fabrication process in a graphene flake are conceivable\cite{Castro,Mariani,Morozov,Stauber}, but also atomic defects in substrates like hBN could induce nanometer-scale strain variations\cite{Anota,Park,Satos}.
%Under compressive strain nanometer-scale strain variations may be intricately mingled with the formation of static ripples.\cite{Castro,Meyer,Fasolino,Braghin,Monteverde} We therefore focus our analysis on applied tensile strain and consider both the effect of thermal fluctuations and of a static Gaussian potential to model defects in a hBN substrate.
%
For both potential sources of nanometer-scale strain variations we observe an intriguing non-monotonic variation of the average atomic displacements with increasing externally applied tensile strain. Arising from the non-linear elastic properties of graphene, this variation allows us to predict experimentally observable signatures of nanometer-scale strain variations in the measurement of the carrier mobility as well as in the measurement of the line-widths of the Raman active G- and 2D-mode.

\section{Methods}

Aiming for system sizes beyond the reach of first-principles approaches we base our molecular modeling study on the empirical long-range carbon bond order potential (LCBOP)\cite{Los}. Among a multitude of force fields for carbon materials\cite{Los,Brenner,Ni,Stuart,vanDuin,Tersoff} LCBOP exhibits a range of features that are critical to the targeted properties. Specifically, it yields sound velocities, a bending rigidity and elastic constants that match the experimental values within $\sim 10\%$.\cite{Los} Of particular relevance for the present study is hereby the faithful representation of the highly nonlinear elastic properties of graphene\cite{Kalosakas,Sloan,Cadelano}. This is illustrated in Fig. \ref{fig1new}, which shows the calculated stress in a graphene sheet as a function of externally applied strain as obtained with density-functional theory (DFT) using the CASTEP package\cite{Clark} (library pseudopotentials, 650 eV cutoff energy, and $(7 \times 7)$ k-point sampling) and treating electronic exchange and correlation at the level of the PBE functional\cite{Perdew}.
%together with the Tkatchenko-Scheffler dispersion-correction scheme\cite{Tkatchenko,McNellis}}.
%%MA Here only PBE was used. As this is free-standing graphene there is no vdW bonding to substrate.

Already at rather low strain values the stress-strain curve deviates from the linear behavior expected for a material with constant Young's modulus of 1.02~TPa \cite{Lee}. Reproducing the findings of Kalosakas {\em et al.}\cite{Kalosakas} the LCBOP data faithfully reproduces this behavior and achieves a match to the first-principles calculations within $\sim 10\%$.

\begin{figure}[!b]
 \begin{center}
     \includegraphics[width=86mm]{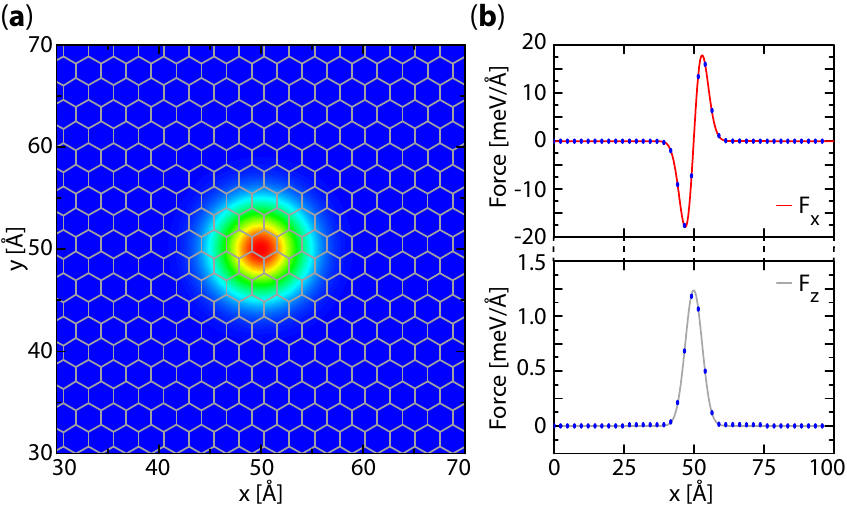}
 \end{center}
 \caption{(Color online) Illustration of the simulation cell and the applied Gaussian potential. The Gaussian potential has a width of three interatomic carbon-carbon bonds and a strength of +90 meV. (a) Top view of the central area of the graphene sheet with all individual C atoms overlayed with a color map of the Gaussian potential that is applied in the center of the sheet. (b) Force induced by the Gaussian potential on the C atoms of an ideal sheet and along a line at $y = 50$\,{\AA} (red = in-plane force, gray = out-of-plane force, lines are superimposed to guide the eye).}
 %%KR Please adapt this figure: Remove panel (c) - it is not necessary and one would need a lot of space to explain properly what is shown here. Please also modify panel (b): Include data points at the position of every C atom along this line. The continuous lines suggest that graphene is modeled as a continuum, which is not correct. - GV: Done.
 \label{fig1}
\end{figure}

All simulations were carried out with the LAMMPS package\cite{Plimpton}. The standard simulation cell contains a rectangular sheet of graphene with a side length $L$ of approximately 10 nm (3680 atoms), corresponding to optimized LCBOP carbon-carbon distances. Periodic boundary conditions are applied and a vacuum layer of 20\,{\AA} separates the graphene sheet from its periodic images. Static nanometer-scale strain variations induced e.g. by lattice imperfections in an underlying hBN substrate are modeled by subjecting the graphene sheet to the influence of an isotropic Gaussian potential\cite{Carillo,Moldovan}. This potential is characterized by three parameters: its width, its strength and its perpendicular distance to the graphene sheet. Based on the discussion above, the width of the Gaussian potential must be much smaller than 100\,nm. We performed simulations varying the width of the potential between one and four times the interatomic carbon-carbon distance in unstrained graphene (1.42 - 5.68\,{\AA}), which results in a strain variation of a few nanometer in size (see below). The conclusions put forward below are identically derived for all widths, which is why we restrict the detailed presentation of the results below to the case of three times the interatomic carbon-carbon distance. Preceding calculations suggest a maximum potential difference induced by point defects in hBN of the order of 100\,meV.\cite{Anota,Park,Satos} We find this confirmed by our own DFT calculations for a nitrogen surface vacancy and a boron interstitial. We correspondingly performed simulations varying the strength of the Gaussian potential in 10\,meV steps between -100\,meV (attractive) and 100\,meV (repulsive). We obtain qualitatively the same results for all cases, and discuss below some selected values for the potential strength. The perpendicular distance of the Gaussian potential to the graphene sheet is fixed at 0.25\,{\AA} to induce out-of-plane displacements in the graphene sheet of similar magnitude as the in-plane displacements over the entire range of applied external strain. By doing so, we obtain both maximum in-plane forces $F_{x,y}$ and out-of-plane forces $F_z$ on the graphene sheet that are typically of the order of several meV/{\AA} (1 meV/{\AA} = 1.6 pN). This is illustrated in Fig. \ref{fig1}.

The effect of thermal fluctuations is modeled through a Nos\'{e}-Hoover thermostat\cite{Nose} with the temperature set to 300\,K. To accommodate the effect of thermal expansion the lateral cell size was equilibrated in 4\,ns molecular dynamics (MD) runs at zero applied pressure and using the velocity-Verlet integration scheme with 0.001\,ps time steps. As the obtained fluctuations in the atomic positions in the graphene sheet are known to scale with the side length of the sheet\cite{Fasolino}, simulations were performed for various side lengths/number of carbon atoms in the cell to obtain the scaling relation between the atomic displacements and side length $L$.

For both the static Gaussian potential and the dynamic thermal fluctuation simulations the external strain is applied by rescaling the lateral simulation cell size. Specifically, we consider the effect of uniaxial strain along the zigzag direction ($x$ direction, cf. Fig. \ref{fig1}) and along the armchair direction ($y$ direction, cf. Fig. \ref{fig1}), as well as the effect of equibiaxial strain ($x$ and $y$ direction). In the uniaxial cases, the simulation cell size was kept fixed in the respective other direction. The applied strain is always  varied from 0\% to 0.5\% in steps of 0.1\% and from 0.5\% to 15\% in steps of 0.5\%.

The central outcome of the simulations for a given external strain are the average atomic displacements and the relative changes in bond length. Distinguishing the average in-plane displacement $d_{\rm ip}$, the average out-of-plane displacement $d_{\rm oop}$, and the average total displacement $d_{\rm tot}$, these are defined as
\begin{align}
d_{\rm ip}  &= \sum_{i=1}^{N} \frac{\sqrt{(x_i - x_{i,0})^2 + (y_i - y_{i,0})^2}}{N},\nonumber\\
d_{\rm oop} &= \sum_{i=1}^{N} \frac{\left|z_i - z_{i,0}\right|}{N},\label{eq1}\\
d_{\rm tot} &= \sum_{i=1}^{N} \frac{\sqrt{(x_i - x_{i,0})^2 + (y_i - y_{i,0})^2 + (z_i - z_{i,0})^2}}{N},\nonumber
\end{align}
where ($x_i,y_i,z_i$) is the position of atom $i$ at the applied Gaussian potential (MM) or at 300 K (MD), ($x_{i,0},y_{i,0},z_{i,0}$) is its equilibrium position at 0 K without applied Gaussian potential, and $N$ is the total number of atoms in the simulation cell.

For the case of the static Gaussian potential, these displacements are evaluated after relaxing the atomic positions under the applied strain until residual forces fell below 0.5 meV/{\AA}. For the thermal fluctuations the displacements are obtained from MD simulations. Starting from the relaxed zero-strain geometry (see above), the system was first equilibrated under the applied strain over 4\,ns in the $(NVT)$ ensemble. The displacements are then obtained as time averages over 6\,ns $(NVE)$ trajectories.

\section{Results}

\subsection{Effect of a static Gaussian potential}

\begin{figure}[!t]
 \begin{center}
 	\includegraphics[width=86mm]{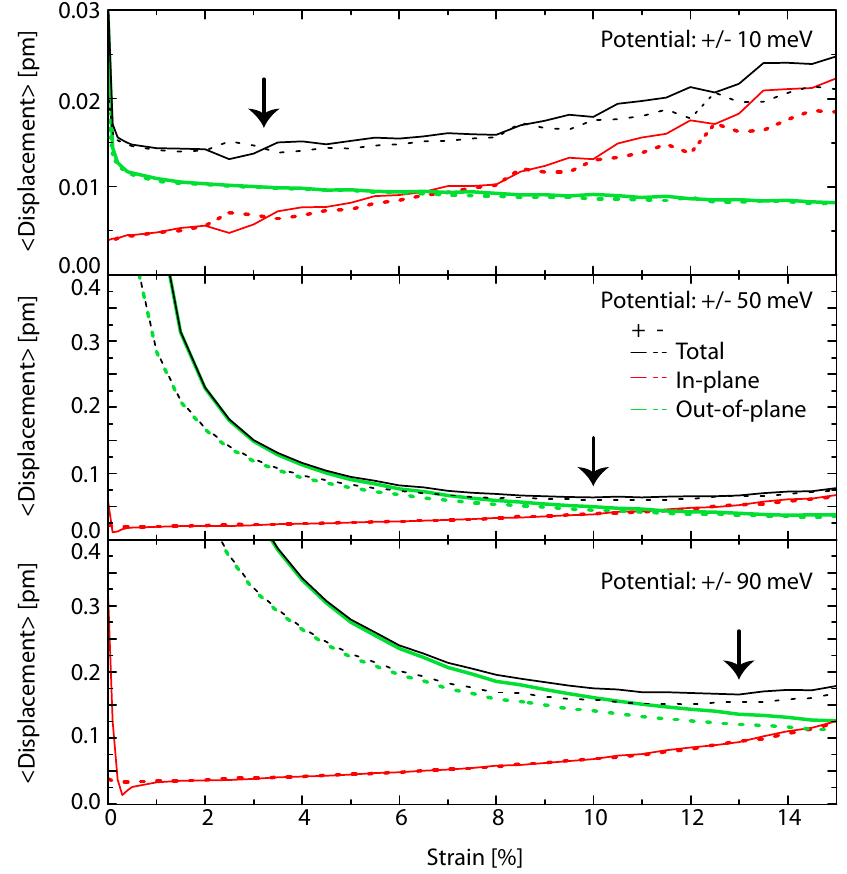}
 \end{center}
 \caption{(Color online) Calculated average displacements under the influence of a static Gaussian potential. Shown as a function of applied equibiaxial tensile strain are the total displacement $d_{\rm tot}$ (black curves), the in-plane displacement $d_{\rm ip}$ (red curves), and the out-of-plane displacement $d_{\rm oop}$ (green curves). The upper panel corresponds to a strength of the Gaussian potential of 10\,meV, the middle panel to 50\,meV, and the lower panel to 90\,meV. In all panels, solid curves correspond to a repulsive potential and dotted curves to an attractive potential. The arrows indicate the minimum in total displacement $d_{\rm tot}$.}
 \label{fig2}
\end{figure}

Figure \ref{fig2} summarizes the obtained average displacements under applied equibiaxial strain and Gaussian potentials of strength $\pm 10$\,meV, $\pm 50$\,meV and $\pm 90$\,meV. For all three potential strengths and irrespective of whether the potential is attractive or repulsive the average out-of-plane displacement $d_{\rm oop}$ decreases monotonically with increasing strain. As the Gaussian potential and thus the force exerted on the graphene sheet is constant, this indicates an increase in the out-of-plane bending stiffness. The average in-plane displacement $d_{\rm ip}$ exhibits exactly the opposite behavior, i.e. it increases with increasing strain. This reflects the decrease of the Young's modulus with increasing strain that was discussed in Section II. Intriguingly, the increase of $d_{\rm ip}$ is in all cases more than linear with strain. The reason for this is that the decrease of Young's modulus is more than linear with the strain (not shown). If one assumes a simple linear elastic model, in which the strong coupling between the in-plane and out-of-plane deformation is ignored, one finds the trivial result that the in-plane displacements should be independent of the strain. As we see a large increase in in-plane deformations, the difference between the in-plane displacement at finite strain and the one at zero strain quantifies how much the in-plane displacements are beyond linear elasticity. In our case, the in-plane displacements are up to a factor 5 larger than those expected from the simple linear elastic model. The opposite behavior of the average out-of-plane and in-plane displacement has the consequence, that the average total displacement $d_{\rm tot}$ reaches a minimum for a certain amount of strain (see arrow). The precise value for this amount depends on the potential strength (as apparent from Fig. \ref{fig2}) and on the potential width (not shown). The stronger and narrower the potential, the higher the strain required to reach the minimum. In fact, for a potential width of only once or twice the size of the interatomic carbon-carbon bond, the minimum is always above 15\% strain for the explored values of potential strength and for the considered, fixed perpendicular distance between the center of the potential and the graphene sheet. This is approximately equal to the breaking strength of graphene \cite{Lee,Kim2,Booth} and will, therefore, not be reachable in experiments.
%Intriguingly, the increase of $d_{\rm ip}$ is in all cases more than linear with strain, i.e. graphene gets softer in the in-plane direction when strained. This reflects the decrease of the Young's modulus with increasing strain that was discussed in Section II.

%%MA I think this discussion fits better here - moved back again...
%%GV [I copied this part of the comment here for the new alinea/discussion]
%%Does it have to do with
%%KR the width/strength of the potential. Nothing is said in this paragraph about the variation of the reported distributions with
%%KR potential width/strength. A corresponding discussion should be added.
Both the average in-plane and out-of-plane displacements depend on the width and the strength of the potential. We observe that the average out-of-plane displacements scale linearly with the strength of the potential, which is explained from the fact that the out-of-plane force scales linearly with the strength of the potential. However, the out-of-plane force is largely independent of the width of the potential, as the distance to the graphene sheet is fixed to 0.25 \AA, which is much smaller than the explored widths in the simulation. Therefore, there is hardly any effect on the average out-of-plane displacements when the width of the potential is changed from one to four interatomic carbon-carbon bonds. The average in-plane displacements, on the other hand, are highly sensitive to both the width and the strength of the potential, since the in-plane force depends on both of these parameters.

\begin{figure}[!t]
 \begin{center}
    \includegraphics[width=86mm]{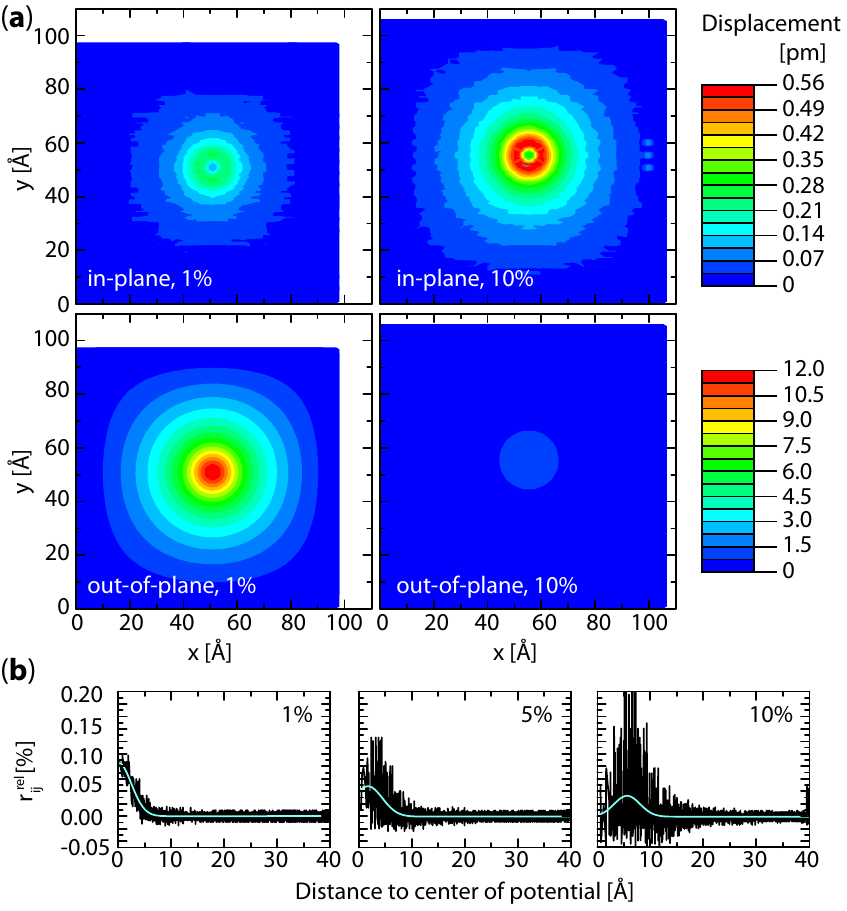}
 \end{center}
 \caption{(Color online) (a) Top view of the graphene sheet depicting the spatial distribution of the atomic displacements under the influence of a Gaussian potential and for equibiaxial tensile strain of 1\% (left panels) and 10\% (right panels). Compared are the in-plane displacements (upper panels) and the out-of-plane displacements (lower panels). The Gaussian potential of strength 90 meV is centered in the middle of the sheet and the discrete displacement data of the individual C atoms in the sheet has been interpolated to a continuous colormap for easier visualization. (b) Relative change in carbon-carbon bond length as a function of the distance to the center where the Gaussian potential is applied. Shown are data for three tensile strain values of 1\%, 5\% and 10\%. The cyan line is a guide to the eye.}
 %%KR Isn't it awkward to have the color legend to the panels of (a) show highest values at the bottom and lowest values at the top? It would feel much more natural to me to have this reversed. - GV: Done.	
 \label{fig3}
\end{figure}

Apart from the average displacements it is also instructive to analyze the spatial distribution of the individual atomic displacements over the graphene sheet.
Figure \ref{fig3}(a) compiles corresponding data for two different equibiaxial strain values.
The overall increase (decrease) of the in-plane (out-of-plane) displacements for the larger strain value (10\%) is again clearly visible.
%%MA "Remarkably" is a bit strong here...
The two types of displacements exhibit qualitatively different spatial distributions, as one would naturally expect from the qualitatively different spatial distribution of the in-plane and out-of-plane forces shown in Fig. \ref{fig1}(b). The
%Remarkably, however, the two types of displacements exhibit qualitatively different spatial distributions.
%As one would naturally expect, the
out-of-plane displacements are largest exactly at the center of the sheet where the potential is applied and decay radially outwards. In contrast, the in-plane displacement distribution has a ring-like form, i.e. maximum displacements are found at some radial distance away from the source of the disturbance. This distance, or in other words the size of the ring, increases with the width of the potential, but is otherwise largely independent of the strength of the potential and the applied strain.

The fact that the in-plane and out-of-plane displacements differ in their spatial distributions, where out-of-plane displacements are largest at the center of the potential while in-plane displacements exhibit a ring-like form, directly translates to the carbon-carbon bond lengths within the sheet. We quantify this via the nearest neighbour vector ${\bf r_{ij}}$ and the relative change in bond length $r_{ij}^{\rm rel}$ between neighbouring carbon atoms $i$ and $j$:
\begin{align}
{\bf r_{ij}} &= \left(x_{j} - x_{i},\, y_{j} - y_{i},\, z_{j} - z_{i}\right),\label{eq1b}\\
r_{ij}^{\rm rel} &= \frac{\left| {\bf r_{ij}} \right| - \left| {\bf r_{ij,0}} \right|}{\left| {\bf r_{ij,0}} \right|}.\nonumber
\end{align}
Here $\left| {\bf r_{ij,0}} \right|$ is determined at 0 K without applied Gaussian potential. Figure \ref{fig3}(b) shows the relative change of this bond length $r_{ij}^{\rm rel}$ as a function of the distance to the center where the potential is applied. At the smallest strain shown (1\%) the variations are dominated by the out-of-plane displacements and largest changes in the bond lengths are correspondingly obtained at the position of the potential. For larger strains, the in-plane displacements gradually take over. Correspondingly, maximum bond length changes are also observed at larger distances from the applied potential. At a strain of 10\% these maximum values are found about 8\,{\AA} away from the potential. The resulting diameter of the nanometer-scale strain variation of about 16\,{\AA} is therewith almost an order of magnitude larger than the width of the potential (4.26~\AA) causing it. This is a consequence of the direct relation between the applied external potential and the resulting deformation potential that arises due to changes in bond lengths \cite{Barraza}.

Even though the maximum bond changes are quite small (less than 0.2\%), it is important to realize that this still results in a significant pseudomagnetic field $B$ \cite{Katsnelson,Pereira2,Masir}:
\begin{align}
B &= \frac{\partial A_y}{\partial x} - \frac{\partial A_x}{\partial y},\label{eq2}\\
A_x + i A_y &= \frac{t_0 \beta}{e v_F} \sum_{ij} r_{ij}^{\rm rel} \exp^{-i {\bf K}\cdot{\bf r_{ij,0}}},\nonumber
\end{align}
%%KR ", which is explained by the much smaller out-of-plane deformation (0.12 nm at 1\% strain) in our calculations."
%%KR The latter sentence is not an explanation. That one obtains smaller fields for smaller deformations is self-evident. The question
%%KR is why we obtain much smaller deformation as Carillo and Moldovan? Do you have an explanation for this?
in which $r_{ij}^{\rm rel}$ and ${\bf r_{ij,0}}$ are defined in Eq. \ref{eq1b}, ${\bf A} = (A_x,\,A_y)$ is the pseudovector potential, $t_0 = 2.8 eV$ is the so-called hopping parameter, $\beta$ is the strained hopping energy modulation factor, $e$ is the electron charge, $v_F$ is the Fermi velocity, and ${\bf K}$ is the high symmetry point at the edge of the Brillouin zone. For 1\%, 5\% and 10\% strain, we obtain maximum pseudomagnetic fields of 60 mT, 110 mT and 160 mT, respectively.
These numbers are much smaller than those reported in previous theoretical investigations \cite{Carillo,Moldovan}.
The reason for this is that the previous work investigated a graphene structure with a much larger ratio between the out-of-plane displacement ($\sim 2.5$ nm) and the width ($\sim 5$ nm) of the nanometer-scale strain variation and furthermore did not allow the atoms to relax in-plane.
As a consequence, the gradient of the relative bond length change $r_{ij}^{\rm rel}$, which is the important quantity for the strength of the pseudomagnetic field, is in our case much smaller, which thus results in a much smaller pseudomagnetic field.
Nevertheless, such small pseudomagnetic fields do affect the uniformity of the pseudomagnetic field in a graphene flake.

\begin{figure}[!t]
 \begin{center}
    \includegraphics[width=86mm]{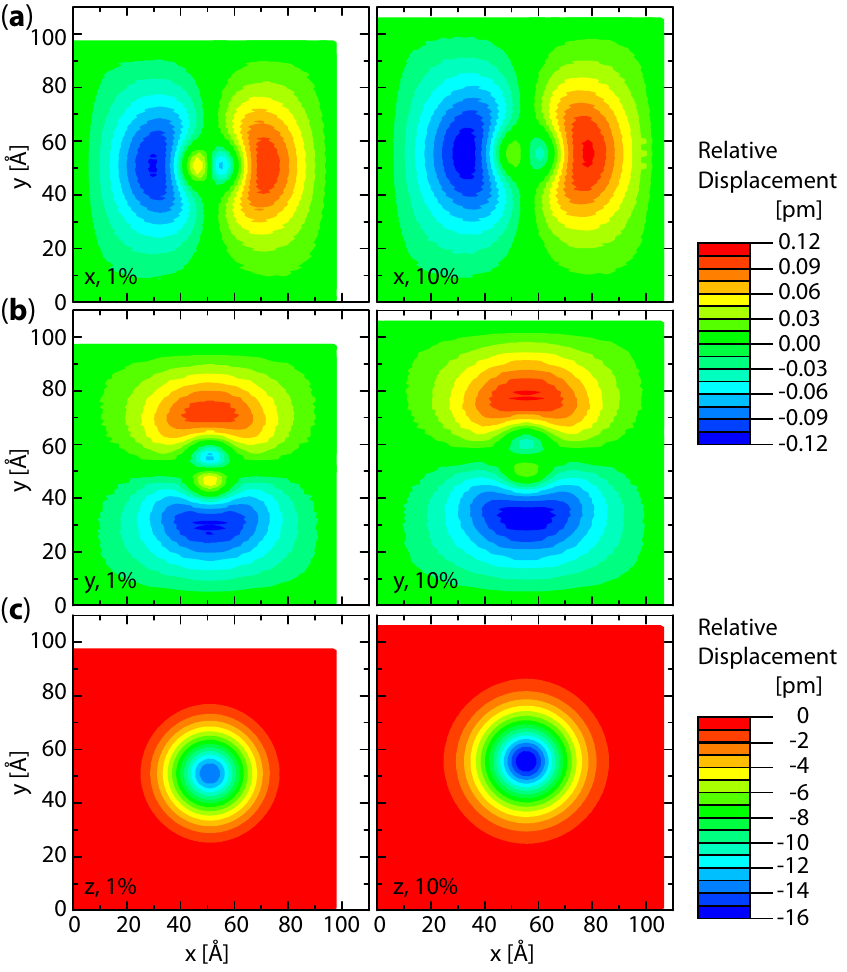}
 \end{center}
 \caption{(Color online) Calculated relative displacements with respect to the atomic positions of the graphene sheet with a Gaussian potential of 90 meV at zero applied strain. The panels show from top to bottom the $x$-, $y$-, and $z$-component at 1\% and 10\% of applied strain. In agreement with Fig. \ref{fig2} and \ref{fig3}, the in-plane displacements increase with applied strain, whereas the out-of-plane displacement decreases.}
 \label{fig7}
\end{figure}

As Gaussian potentials \cite{Carillo,Moldovan} or bubbles with different geometries \cite{Bahamon} induce inhomogeneous strain distributions in the graphene sheet with regions of compression, elongation and unaffected bond length, it is imperative to consider the change in displacements in the nanometer scale strain variation itself and thus take the atomic positions of the graphene sheet with the Gaussian potential at zero applied external strain as reference position. Figure \ref{fig7} compiles the result for a Gaussian potential with a strength of 90 meV. The displacement in $x$- and $y$-direction of the atoms, that together form the in-plane displacement, are enhanced when going from 1\% to 10\% of applied tensile strain. This is fully consistent with the results in Figs. \ref{fig2} and \ref{fig3}. In Figs \ref{fig7}a and \ref{fig7}b, the outer red and blue lobes indicate that the atoms are pushed radially outwards with respect to the position where the Gaussian potential is applied. In contrast, the inner lobes represent a region of compression and is a result of the strong coupling between the in-plane and out-of-plane displacement. The out-of-plane displacement $z$ keeps decreasing with applied tensile strain and reaches almost the out-of-plane displacement of the reference configuration at 10\% applied strain.

\begin{figure}[!t]
 \begin{center}
    \includegraphics[width=86mm]{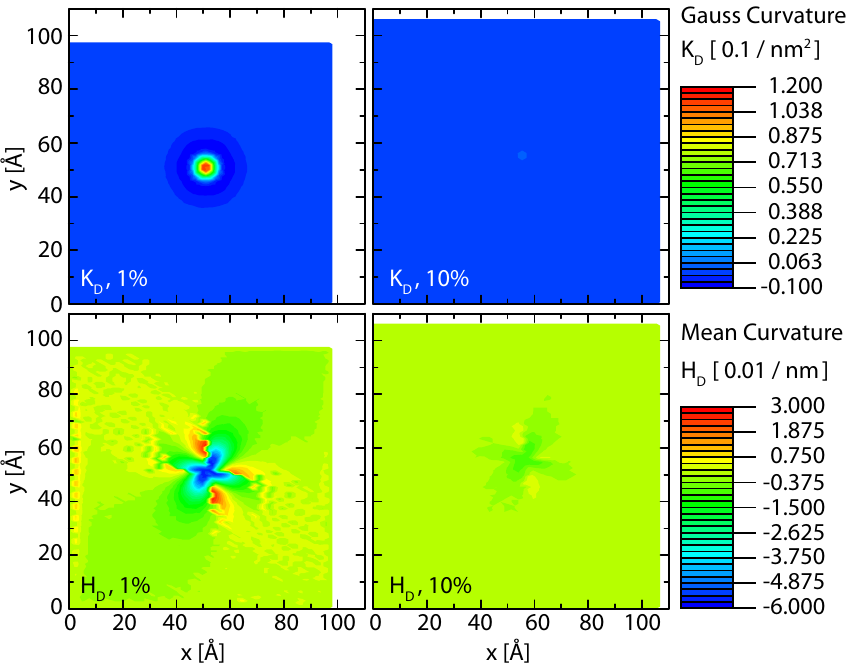}
 \end{center}
 \caption{(Color online) Calculated discrete Gaussian curvature $K_{\rm D}$ and discrete mean curvature $H_{\rm D}$ for a Gaussian potential of 90 meV at 1\% and 10\% of applied strain. Both quantities decrease with at least one order of magnitude when increasing the applied strain from 1\% to 10\%.}
 \label{fig6}
\end{figure}

Finally, we consider the change in local geometry of the two-dimensional surface \cite{Sanjuan}, which is characterized by the discrete Gaussian curvature $K_{\rm D}$ and the discrete mean curvature $H_{\rm D}$. The former quantifies the local bending of the surface, whereas the latter measures the relative orientation of edges and normal vectors along a closed path. Both are zero in the case of a flat surface. Figure \ref{fig6} shows both quantities for a Gaussian potential of 90 meV at an applied strain of 1\% and 10\%. Both, the discrete Gaussian curvature $K_{\rm D}$ and discrete mean curvature $H_{\rm D}$ show that the graphene becomes flatter with increasing strain, as both quantities decrease at least one order of magnitude. This is consistent with the observed decrease in out-of-plane displacements.

\begin{figure}[!t]
 \begin{center}
    \includegraphics[width=86mm]{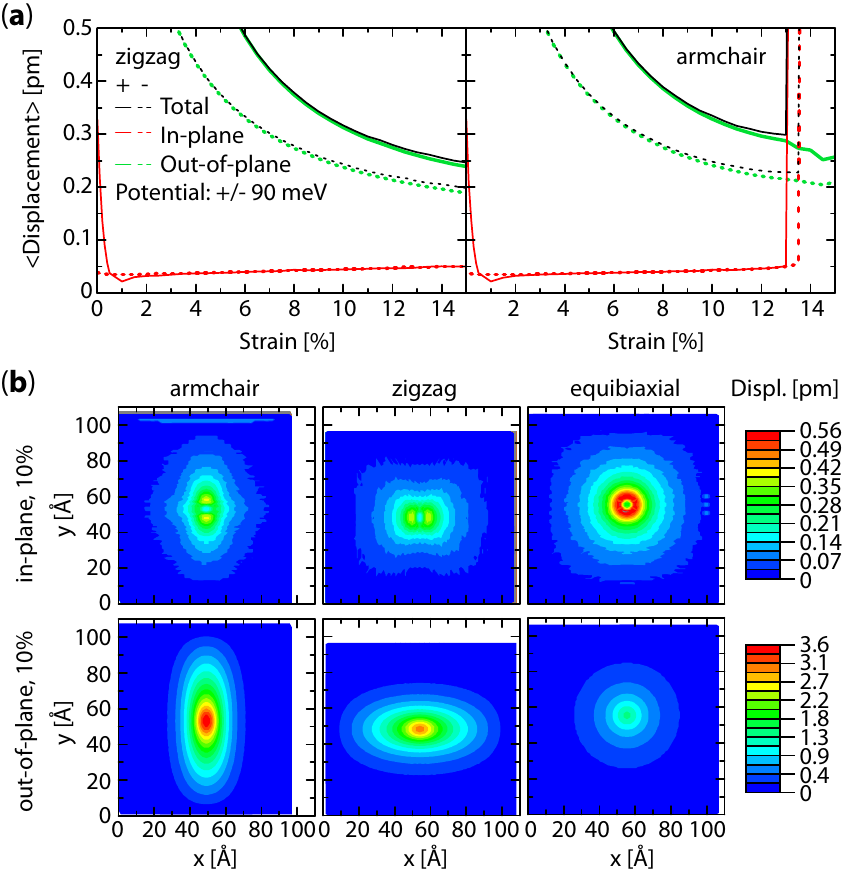}
 \end{center}
 \caption{(Color online) (a) Calculated average displacements under the influence of a static Gaussian potential of strength 90\,meV. Shown as a function of applied uniaxial tensile strain along the zigzag (left panel) and armchair (right panel) direction are the total displacement $d_{\rm tot}$ (black curves), the in-plane displacement $d_{\rm ip}$ (red curves), and the out-of-plane displacement $d_{\rm oop}$ (green curves). In both panels, solid curves correspond to a repulsive potential and dotted curves to an attractive potential. (b) Top views of the graphene sheet depicting the corresponding spatial distribution of the atomic displacements under 10\% strain. Compared are the in-plane displacements (upper panels) and the out-of-plane displacements (lower panels) for uniaxial strain along the armchair (left panels) and zigzag (middle panels) direction, as well as equibiaxial strain (right panels).}
 %%KR Isn't it awkward to have the color legend to the panels of (b) show highest values at the bottom and lowest values at the top? It would feel much more natural to me to have this reversed. - GV: Done.	
 \label{fig4}
\end{figure}

In the simulations with uniaxial strain, both along the zigzag and the armchair direction, we qualitatively obtain equivalent findings as those just summarized for the equibiaxial strain case. This is visible in the average displacements compiled in Fig. \ref{fig4}(a).
Also here, we find the decrease (increase) of the average out-of-plane (in-plane) displacements with strain. However, since the ratio of the out-of-plane to in-plane displacements is much larger for uniaxial strain, a minimum in the total average displacement is not found (within the range of strain values below the breaking point of graphene) for the investigated parameters of the Gaussian potential. The breaking of the graphene sheet can be observed as a sudden rise in the in-plane displacements at a strain around 13\% along the armchair direction in Fig. \ref{fig4}(a).

Another important difference between equibiaxial and uniaxial strain is that the lower symmetry of the uniaxial strain leads to an anisotropy in the underlying spatial distributions, which are summarized in Fig. \ref{fig4}(b).
For both straining directions it is seen that the spatial distributions of the in-plane and out-of-plane displacements are elongated along the direction of the applied strain.
Considering the range of the observed displacements, it is obvious that the displacement of the atoms close to the center of the potential affects a number of neighbor atoms extending far beyond nearest neighbors and even outside of the area of the graphene sheet where the potential has any appreciable strength.
To explain the observed spatial distributions of the displacements, we will assume that the range of the displacements, for a given magnitude of displacement of the central atoms, corresponds to a fixed number of carbon atoms along any direction outwards from the center of the potential.
Thus, the elongated carbon-carbon bonds along the straining direction should result in a larger range of the displacements.
For the in-plane displacements, where the central atoms experience a similar magnitude of displacement in the uniaxial and equibiaxial case, it is indeed seen that the range of the in-plane displacements along the straining direction in both the armchair and zigzag case closely matches the range observed along the corresponding straining direction in
the equibiaxial case.
For the out-of-plane displacements it is seen that the magnitude of the displacements are much larger in the uniaxial case, which indicates that the out-of-plane bending stiffness is increased less for uniaxial strain than for biaxial strain.
As expected, the increased magnitude of the displacements leads to an increased range of the displacements both perpendicular to and along the straining direction.
Thus, in this case the range of the displacements perpendicular to the straining direction is comparable to the equibiaxial case,\linebreak
\onecolumngrid

\begin{figure}[!t]
 \begin{center}
    \includegraphics[width=\textwidth]{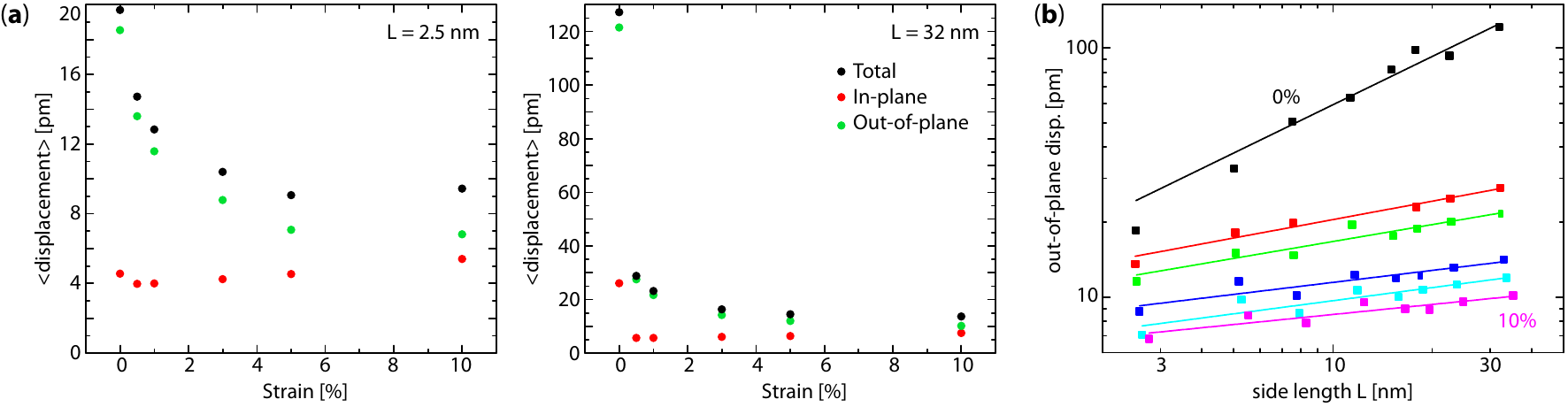}
 \end{center}
 \caption{Calculated time-average displacements as a function of strain from MD simulations at 300 K. Results are shown for a side length $L$ of 2.5 nm (left panel) and 32 nm (middle panel). In agreement with the MM simulations, $d_{\rm oop}$ decreases with increasing strain, whereas $d_{\rm ip}$ increases. The right panel shows how the average out-of-plane displacement scales with the side length $L$ for different values of strain: 0\% (black), 0.5\% (red), 1\% (green), 3\% (blue), 5\% (cyan), 10\% (magenta). The slope of the fitted lines yields the critical exponent of membrane theory, see text.}
 \label{fig5}
\end{figure}
\twocolumngrid
\noindent
while the range of the displacements along the straining direction are much larger than the equibiaxial case.
Note that as the out-of-plane bending is always perpendicular to the straining direction, the out-of-plane bending stiffness is identical for all atoms independently of the straining direction.
The in-plane stretching mode, however, is expected to soften along the straining direction as a consequence of the decrease of Young's modulus with increasing strain.
Thus, a part of the observed increase in the in-plane displacements along the straining direction could also be explained by such strain-induced softening.

\subsection{Effect of thermal fluctuations}

% finally introduce the MD part
% total displacement is reduced (see configs)
We now proceed to the effect of thermal fluctuations as determined by the MD simulations at 300 K. The left and center panel in Fig. \ref{fig5} show the time-averaged values of the displacements obtained with Eq. \ref{eq1} as a function of applied strain for a side length $L$ of 2.5 and 32 nm respectively. In both panels, we observe that the out-of-plane displacement $d_{\rm oop}$ (green dots) decreases with increasing strain, whereas the in-plane displacement $d_{\rm ip}$ (red dots) increases. As a consequence, the total displacement $d_{\rm tot}$ (black dots) reaches a minimum for a certain strain value.
These results are thus in perfect agreement with the results of the MM simulations we discussed before.
Overall, the results of our MM and MD simulations suggest that the out-of-plane displacements $d_{\rm oop}$ always decrease with increasing strain and that in-plane variations always increase with increasing strain. For this, it is not important what the origin of the strain variation is.

\begin{table}[!b]
\begin{ruledtabular}
\begin{tabular}{ccc}
  strain [\%] & $\xi$ & error [\%]\\
  \hline
  0 & 0.64 & 10\\
  0.5 & 0.25 & 8\\
  1 & 0.23 & 14\\
  3 & 0.16 & 18\\
  5 & 0.18 & 18\\
  10 & 0.14 & 19\\
\end{tabular}
\end{ruledtabular}
\caption{Critical exponent $\xi$ and its error as a function of the applied strain. The exponent $\xi$ at 0\% strain is 0.64, which is in perfect agreement with literature \cite{Kownacki,Doussal,Gazit,Los,Zakharchenko}.}
\label{tab1}
\end{table}

% plot fluctuation as a function of system size: compare with literature
2D crystals, such as graphene, were thought to be thermodynamically unstable due to their extremely small thickness \cite{Mermim}. In particular, the ratio between its effective thickness, which is defined as the average out-of-plane fluctuations, and the side length of a rectangular 2D crystal $L$, diverges if $L\rightarrow\infty$. Therefore, it was a big surprise that suspended graphene devices could be made and were stable at room temperature. The reason why graphene is stable is the strong coupling between its bending and stretching modes \cite{Doussal}. This coupling is well described in membrane theory by a so-called critical exponent $\xi$ \cite{Wilson,Ma,Patashinskii}. We determine the dependence of this critical exponent on the applied strain, by determining the average out-of-plane displacement $d_{\rm oop}$ as a function of the side length $L$ for different values of strain, cf. Fig. \ref{fig5}. The resulting values are compiled in Table \ref{tab1}. The exponent $\xi$ of 0.64 at zero strain is in good agreement with analytical work done on the statistical mechanics of flexible membranes \cite{Kownacki}, self-consistent screening approximation \cite{Doussal,Gazit}, and MD simulations reported in literature \cite{Los,Zakharchenko}, which give an exponent of 0.58.
% make comparison at finite strain values here!
For higher strain values, the critical exponent decreases and already reaches $\xi =$ 0.25 at 0.5\% of strain. This result indicates that even a small amount of strain stabilizes graphene ($\xi < 0.5$), e.g. it remains flat in the sense that the average out-of-plane displacement remains much smaller than the side length $L$ even if $L\rightarrow\infty$.

\section{Summary and Conclusions}

We presented a molecular modeling study analyzing the effect of externally applied tensile strain on nanometer-scale strain variations of graphene. As source for such nanometer-scale strain variations we considered thermal fluctuations and a static Gaussian potential to model lattice imperfections in an underlying substrate, as the exact physical origins of these nanometer-scale strain variations are hard to capture. For both sources and for both uniaxial and equibiaxial strain the central outcome of our simulations is a decrease of the out-of-plane atomic displacements and a surprising increase of the in-plane atomic displacements with increasing strain. Details of both variations are beyond the expectations of linear elasticity theory, and reflect a strong increase of the bending rigidity and the known decrease of graphene's Young's modulus with strain, respectively. In conjunction both trends lead generally to a non-monotonic change in the total average displacement with strain. This displacement first drops with applied strain and then starts to rise again.

This finding suggests an intriguing opportunity to further investigate the influence of nanometer-scale strain variations as the major mobility limitation to graphene. If so, then the mobility of suspended graphene should initially increase with applied tensile strain, only to decrease again after a critical amount of strain. In contrast, we expect the confining potential of sandwich structures of the type hBN/graphene/hBN to largely suppress out-of-plane displacements. Leaving only in-plain strain variations, the mobility should in this case consistently be reduced upon application of tensile strain.
The critical role and variation of the in-plane displacements could thereby be further validated through (tip-enhanced) Raman spectroscopy\cite{Ferrari,Basko,Beams}. Both the Raman active G-mode and 2D-mode are in-plane modes and are therefore expected to be sensitive to in-plane strain variations. Assuming the width of these modes to reflect the amount or strength of nanometer-scale strain variations, our simulations suggest an increase of this width with increasing tensile strain. The known splitting of the G-mode in the case of uniaxial strain \cite{Mohiuddin,Ni2,Huang2} could thereby represent a complication. However, such a splitting has not been observed for the 2D-mode and does also not occur for the G-mode in the case of equibiaxial strain \cite{Ding,Zabel,Pan,Jie}. As such, externally applied strain appears as a highly suitable experimental knob\cite{Garza,Garza2,Zhang4} to further understand the role of nanometer-scale strain variations in determining the material's properties of high-quality graphene.

The presented insights may be also highly relevant for other two-dimensional crystals with non-linear Young's modulus. As these materials have an unique electromechanical coupling analogous to graphene, a detailed understanding of this interplay is crucial for future applications of these two-dimensional crystals as well.
\\

\section{Acknowledgements}

S.E.H. and M.A. acknowledge funding from the Alexander von Humboldt foundation. G.J.V. and C.S. acknowledge funding from the ERC (GA-Nr. 280140) and the EU Flagship-Graphene (contract no. NECT-ICT-604391).

\end{document}